\documentclass[12pt, draftclsnofoot, onecolumn]{IEEEtran}

\usepackage{mathptmx}

\usepackage{array}
\usepackage{latexsym}
\usepackage{stmaryrd}
\usepackage{amsmath}
\usepackage{amssymb}
\usepackage{amsfonts}
\usepackage{color}
\usepackage{epsfig}
\usepackage{boxedminipage}
\usepackage{times}
\usepackage{verbatim}
\usepackage{url}
\usepackage{xspace}
\usepackage{algorithm}
\usepackage{algpseudocode} 
\usepackage{paralist}
\usepackage{subfig}
\newcommand{\rom}[1]{\uppercase\expandafter{\romannumeral #1\relax}}

\begin{document}
\title{Analysis of PCA Algorithms in Distributed Environments}

\author{ 
Tarek Elgamal and Mohamed Hefeeda \\
Qatar Computing Research Institute \\
Qatar Foundation \\
Doha, Qatar  \\
Technical Report \\
20 April 2015
 }
\maketitle

\begin{abstract}
Classical machine learning algorithms often face scalability bottlenecks when they are applied to large-scale data. 
Such algorithms were designed to work with small data that is assumed to fit in
the memory of one machine. In this report, we analyze different methods for
computing an important machine learing algorithm, namely Principal Component
Analysis (PCA), and we comment on its limitations in supporting large datasets.
The methods are analyzed and compared across two important metrics: time complexity
and communication complexity. We consider the worst-case scenarios for both
metrics, and we identify the software libraries that implement each method. The analysis in this report helps researchers and engineers in (i) understanding
the main bottlenecks for scalability in different PCA algorithms, (ii) choosing
the most appropriate method and software library for a given application and
data set characteristics, and (iii) designing new scalable PCA algorithms.
%This report provides detailed, step-by-step, analysis of various 
%PCA methods in the literature. 

\end{abstract}

\section{Introduction} \label{sec:intro}

Enormous amounts of data are being generated every day from social networks,
sensors and web sites. This data often has significant value for business,
science, government, and society. Owners of this data strive to extract useful
information out of it, often by applying machine learning algorithms.
Distributed machine learning algorithms are needed because of the large data
volumes and the need to obtain fast results.
Distributed machine learning algorithms, however, introduce a new set of
challenges. For example, during the distributed execution of a machine learning
algorithm, processing nodes may need to exchange data among each other, which we call intermediate data. If not carefully managed, this intermediate data may actually become the main bottleneck for scaling machine learning algorithms, regardless of the available number of computing nodes.

In this report, we focus on analyzing the methods for computing principal
component analysis (PCA). We analyze all methods across two important metrics: time complexity and communication complexity. We consider the worst-case scenarios for both metrics. The time complexity is the upper bound on the number of computational steps needed by the algorithm to terminate. The communication complexity is the worst-case total size of the intermediate data. In addition, during our analysis, we identify the methods implemented in common libraries such as
Mahout, MLlib, and ScaLAPACK.  Mahout \cite{mahout} is a collection of machine
learning algorithms implemented on Hadoop MapReduce. MLlib \cite{mllib} is a Spark 
implementation of some common machine learning algorithms. ScaLAPACK
\cite{scalapack} is a library of linear algebra algorithms implemented for 
parallel distributed memory machines. 

The analysis in this report helps researchers and engineers in (i) understanding
the main bottlenecks for scalability in different PCA algorithms, (ii) choosing
the most appropriate method and software library for a given application and
data set characteristics, and (iii) designing new scalable PCA algorithms: e.g.,
the work in \cite{spca}. To the best of our knowledge, such a rigorous analysis was
never done before, and it is crucial for selecting the proper PCA method for
different environments and datasets. 

The notation we use in this report is mostly consistent with Matlab's programing language. Variable names, including matrices, are composed of one or more letters. Multiplication is indicated with a star ($*$) between the
variables. $M'$ and $M^{-1}$ are the transpose and inverse of matrix $M$,
respectively. $I$ is the identity matrix. Furthermore, $M_i$ denotes row $i$ of
matrix $M$. We use $M_i^j$ to refer to the $j$th element of vector $M_i$.

The rest of this report is organized as follows. In
Section~\ref{sec:cost_model}, we present the details of our distributed
execution cost model that we use to analyze different PCA methods. In
Section~\ref{sec:eigen-decomp}, we analyze the first method for computing PCA which is based on eigenvalue decomposition of the
covariance matrix. In section~\ref{sec:svd}, we analyze the methods for
computing PCA that are based on Singular Value Decomposition (SVD). In Section~\ref{sec:ssvd}, we analyze a PCA method based on a variant
of SVD called Stochastic SVD. In Section~\ref{sec:ppca}, we analyze a PCA method
based on the Probalistic Principal Component Analysis algorithm (PPCA).
Section~\ref{sec:summary} presents a summary of the analysis, and
Section~\ref{sec:conc} concludes the report.

\section{Distributed Execution Cost Model} \label{sec:cost_model}
We analyze all methods across two important metrics: 
time complexity and communication complexity. We consider the worst-case 
scenarios for both metrics. The time complexity is the upper bound on the 
number of computational steps needed by the algorithm to terminate. 
Some PCA algorithms run multiple iterations of the same code, where each iteration improves 
the accuracy of its predecessor by starting from a better initial state. The time complexity that we present is 
for a single iteration, as the number of iterations is typically bounded by a
small constant.

During the distributed execution of a PCA algorithm, processing nodes may need to exchange data 
among each other, which we call intermediate data. The worst-case total size 
of the intermediate data is considered as the communication complexity.
We note that most PCA algorithms work in multiple synchronous phases, 
and the intermediate data is exchanged at the end of each phase. That is, 
a phase must wait for the entire intermediate data produced by its predecessor
phase to be received before its execution starts. Therefore, a large amount of
intermediate data will introduce delays and increase the total execution time of
the PCA algorithm, and hence the intermediate data can become a major bottleneck. 
The exact delay will depend on the cluster hardware (network topology, link speed, I/O 
speed, etc.) as well as the software platform used to manage the cluster and run the PCA code. 
Some software platforms, e.g., Hadoop/MapReduce, exchange intermediate 
data through the distributed storage system, while others, e.g., Spark,  
exchange data through shared virtual memory. For our analysis of communication 
complexity to be general, we consider the total number of bytes that need to be 
exchanged, and we abstract away the details of the underlying hardware/software architecture.

%\section{Detailed Analysis of PCA Methods} \label{sec:analysis} 

% XXXX Tarek: Please revise carefully and use the same notations as 
% in the main paper. Maintain consistency. Remove any redundancy. XXXX
% 
% XXX you need to show the time and comm complexities for every method as we mentioned in the paper. 
%In this section, we analyze different methods for computing the principal
%components for a given dataset represented as matrix.

\section{Eigenvalue Decomposition of the Covariance
matrix}\label{sec:eigen-decomp}

A simple way of computing PCA of a matrix $A$ is to compute
the eigenvalue decompositon of its covariance matrix. Given $N \times D$ matrix
$A$, a target dimensionality $d$, the algorithm computes an
$N \times d$ matrix $V$ such that the columns of $V$ are the principal
components of $A$. The algorithm for computing the eigenvalue decomposition of
the covariance matrix can be summarized in the following steps:
\begin{itemize}
   \item{\bf Step1: Compute the mean-centered matrix $A_c$}

   The mean-centered matrix $A_c$ is computed by subtracting the vector of
   all the column means of matrix $A$ from each row of $A$.
%   Since the matrix needs to be centered before eigenvalue decomposition. the
%   first step is to compute the column mean of {\it A} and subtract it from the
%   whole matrix. The centered matrix {\it $A_m$} is generated as follows:
%   
%  	 $A_m= A - \frac{1}{N*D} \sum_{i=1}^{N} \sum_{j=1}^{D} A_i^j$
   \item{\bf Step 2: Compute the covariance matrix $Cv=A_c^{'}*A_c$}
   
   The covariance matrix is computed by multiplying the mean-centered matrix
   $A_c$ with its transpose. This is a computationally intensive step 
   because it requires multiplying two matrices where typically none of them
   can fit in memory.
   
   \item{\bf Step 3: Compute the eigenvalue decomposition of $Cv$}
   
   Eigenvalues $\lambda_i$ and eigenvectors $v_i$ satisfy the relation:
   $$Cv*\lambda_i=\lambda_i*v_i,$$
   where  $\lambda_i$ is the $i^{th}$ eigenvalue and $v_i$ is its
   corresponding eigenvector.
   Putting the above formula in the matrix form:
    $$Cv*\Lambda=\Lambda*V,$$
    where $\Lambda$ is a diagonal matrix whose elements are the eigenvalues
    of the covariance matrix $Cv$, and $V$ is the matrix whose columns are the
    corresponding eigenvectors.The eigenvalue decomposition of the matrix $Cv$
    is given by:
    $$Cv=\Lambda*V*\Lambda^{-1}.$$
    Several algorithms are used to perform such factorization~\cite{eigen}; they
    include the QZ algorithm~\cite{qz}, and Cholesky factorization~\cite{chol}
    when the matrix is symmetric.
    % householder transformations, Givens rotations
    %(will be described in details later), power iterations and QR algorithm.
    %% The asymptotic complexity of such algorithms are less than or equal the
    % complexity of the previous step.
    	
     \item{\bf Step 4: Get the principal components }
   
   The principal components of the input matrix {\it A} are the eigenvectors
   associated with the largest eigenvalues. Since the eigenvalues in the
   diagonal matrix $\Lambda$ are sorted in a decreasing order, then the
   first $d$ vectors in the matrix $V$ are the principal components of $A$: $V=
   (v_1,v_2,\ldots,v_d)$.
 
   \end{itemize}
 
{\bf Time Complexity:}
The algorithm has two computationally intensive steps: computing the covariance
matrix, and computing the eigenvalue decomposition of the covariance matrix. The
computational complexity of the covariance matrix computations is
$O(ND \times min(N,D))$ which is a result of multiplying two matrices of size $D
\times N$ and $N \times D$, respectively. The other computationally intensive
computation is the eigenvalue decomposition. The work in~\cite{eigen} analyzes
the computational complexity of several algorithms for eigenvalue decomposition.
The worst case complexity of such algorithms is $O(D^3)$ for a matrix of size
$D \times D$. Therefore the overall complexity is $O(ND \times min(N,D) +
D^3)$.

{\bf Communication Complexity:}
In addition to the computational cost, the covariance matrix is large and dense
($D\times D$) which incurs substantial communication cost $O(D^2)$, making the
method not suitable for large datasets.

{\bf Additional Notes:}
The algorithm described in this section is implemented in RscaLAPACK. RscaLAPACK
is an add-on package for the R programming language. R is an open-source
prgoramming language used extensively in statistics and data mining tasks. RscaLAPACK provides a scalable PCA
implementation that can be called from an R program and uses the parallel
linear algebra routines implemented in the ScaLAPACK library. It is worth mentioning that RscaLAPACK includes another scalable
PCA implementation that uses singular value decocmposition (SVD).
RscaLAPACK documentation mentions that the latter is the prefered method for
computing PCA. In the next sections we describe the SVD-based algorithms
including the one implemented in RscaLAPACK.

\section{Singular Value Decomposition (SVD)}\label{sec:svd}
Another way of computing PCA is done by a singular value
decomposition (SVD) of a mean-centered matrix. SVD  decomposes matrix $A$
into three matrices $U$, $\Sigma$ and $V$. If $A$ is mean-centered (i.e.,
subtracted by the column mean of $A$), $V$ gives the principal components of
matrix $A$.

Several methods have been proposed to compute the singular value decomposition
of a matrix. Section \ref{dense} describes early methods for computing SVD for
dense matrices.
Such methods have evolved over time and they are currently being used by some
libraries and frameworks. Section \ref{sparse} describes variants of
SVD algorithms that are optimized for sparse matrices. Such
algorithms offer substantial speedup and efficient memory usage when
matrices are sparse. Section \ref{sec:ssvd} explains an approximate singular
value decompostion algorithm, referred to as Stochastic SVD (SSVD). The algorithm
uses random sampling techniques to compute SVD and provides faster and more
scalable means of SVD computation. The algorithm is implemented in the popular
Mahout/MapReduce machine learning library.

\begin{figure}[h]
	\centering
	%\vspace{1.25in}
	\epsfig{figure=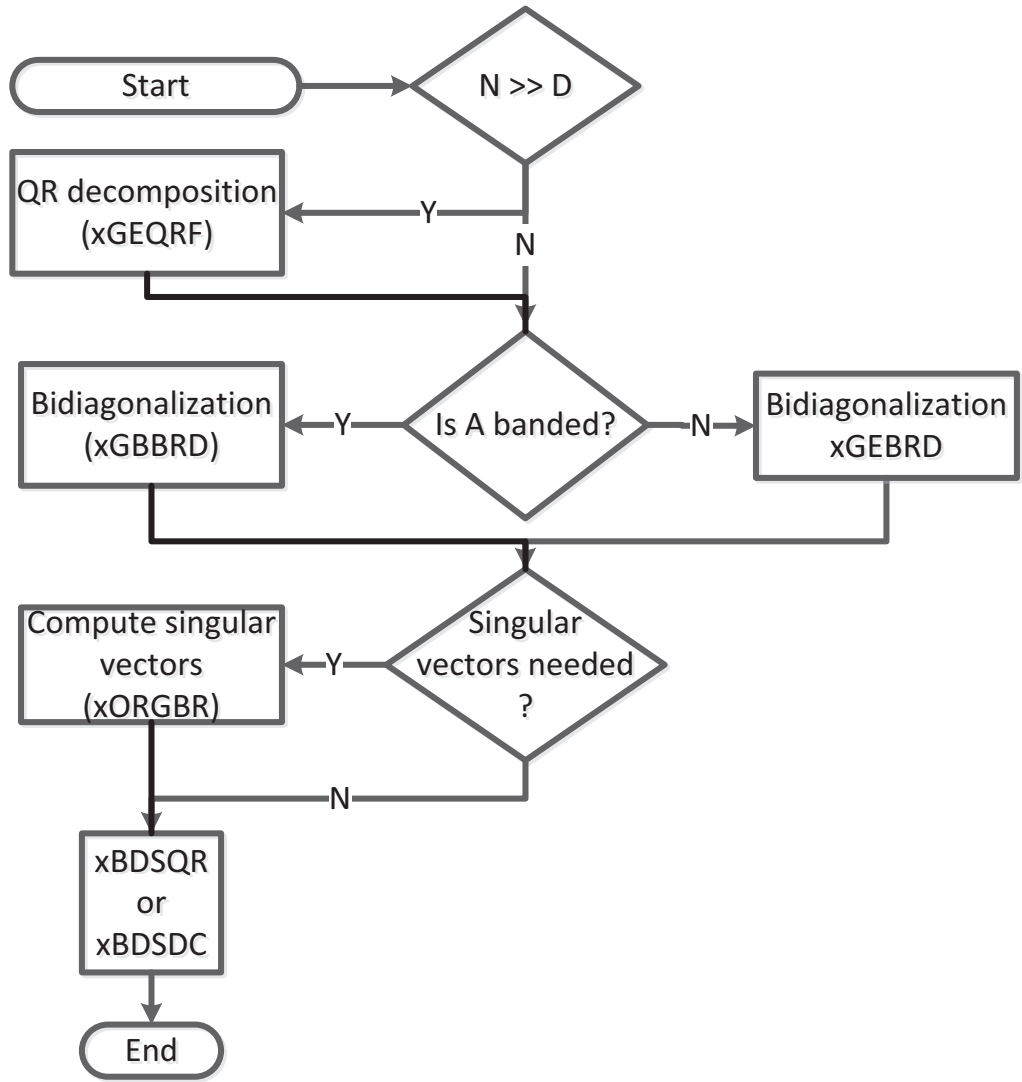,
	height=0.7\textheight}
	%\captionsetup{justification=centering,margin=2cm}
	\caption{Flowchart of the singular value
	decomposition (SVD) algorithm in LAPACK/ScaLAPACK.}
	\label{fig:flowchart_svd}
	\end{figure}

\subsection{Dense Singular Value Decomposition}\label{dense}

Golub and Kahan (1965) introduced a two-step approach for computing
SVD for a given $N \times D$ matrix $A$. In the first step, $A$ is
reduced to a bidiagonal matrix $B$. The second step finds SVD of the bidiagonal
matrix $B$.
Demmel and Kahan \cite{demmel_kahan} extended it to a three-step
procedure. They added a QR decomposition step before the bidiagonalization.
Thus, bidiagonalization is performed for the triangular matrix $R$ instead of
matrix $A$. The algorithm described in this section is implemented in
RscaLAPACK. Figure~\ref{fig:flowchart_svd} shows the flowchart of it. The
figure shows the names of the LAPACK routines used to perform each step. In the rest of the section, we analyze the extended algorithm
described \cite{demmel_kahan}, which can be summarized as:
%The figure shows a dashed rectangle around the branch that will be executed if
%we use this method for large-scale datasets. Morever, 
	
\begin{itemize}
  \item{\bf Step 1: Compute QR decomposition of $A$}
  
  When $N \gg D$, it is more efficient to first perform a QR decomposition
  of $A$. The decomposition reduces the  matrix $A$ into an orthonormal
  matrix $Q$ and a triangular matrix $R$. SVD is then performed more
  efficiently on the triangular matrix $R$ instead of $A$. Since $A=Q*R ,$ and
  $R=U*\Sigma*V$ (described in Step 2), then the SVD of matrix $A$ is given
  by: $$A=Q*U*\Sigma*V.$$

  \item {\bf Step 2: Reduce $R$ to bidiagonal matrix $B$}
  
  As mentioned in the previous step, SVD is performed on the triangular matrix
  $R$. This is done in two steps; 1) Computing bidagonal matrix $B$ and
  2) computing the SVD of $B$. In this step $R$ is reduced to a bidiagonal
  matrix $B$ using a series of Givens rotations as described in \cite{givens}.
  In each iteration, $R$ is multiplied by a rotation matrix $G_i$. Each rotation
  converts one of the elements of $R$ to zero. The series of rotations continues
  until matrix R is converted to a bidiagonal matrix $B$. The output of this
  step is:  
  $$R=U_1*B*V_1,$$
  	 
  	 where each of $U_1$ and $V_1$ is a series of rotations and
  	 $B$ is the bidiagonal matrix computed in this step. $U_1$ and $V_1$ are
  	 defined by: $$U_1=\prod_i G_i,\quad   V_1=\prod_j G_j,$$
  	 $$i \bigcup j= {1,2,3,\ldots , r}.\quad   i \neq j,$$ where $r$ is the is
  	 the total number of rotations required  to convert matrix $R$ to the
  	 bidiagonal matrix $B$.

  \item {\bf Step 3: Compute singular value decompostion (SVD) of the bidiagonal
  matrix $B$}
  
In this step, the QR iteration algorithm is used to compute the SVD of the
bidiagonal matrix $B$. To avoid confusion, it should be noted that the QR
iteration algorithm is different from QR decomposition that is applied in Step
1. The QR iteration algorithm described in this step computes SVD of a matrix,
unlike the one described in Step 1 that decomposes a matrix into two
matrices $Q$ and $R$. 

Before describing how QR iteration algorithm is applied
for bidigonal matrices, we describe how the algorithm works for an arbitrary
real matrix O that is not necessarily bidiagonal. For iteration i, QR
decomposition of matrix O is computed as $O_i = Q*R$, where $Q$ is an
orthonormal matrix and $R$ is upper triangular matrix with positive diagonal.
In the next iteration, $O_{i+1}$ is set to $O_{i+1} = R*Q$. Since matrix $Q$ is
orthornormal (i.e. $Q^{'}*Q=I$) then:
$$O_{i+1} = R*Q = Q^{'}*Q*R*Q = Q^{'}*O_i*Q.$$ 
From the previous equation, it is shown that $O_i$ and $O_{i+1}$ are {\it
similar} matrices. Hence, they have the same eigenvalues. Two
matrices $A$ and $B$ are similar if $B = P^{-1}*A*P$ for some invertible
matrix P. Moreover, it is proved in \cite{golub_kahan_reinsch} that as $i
\rightarrow \infty$, $O_i$ converges to a triangular matrix with the eigenvalues
on the diagonal.
This algorithm may be applied to the bidiagonal singular value problem as
follows \cite{bidiagonal}. 
Let $B_0$ be our initial bidiagonal matrix. Given $B_i$, compute the
QR decomposition $B_i*B_i^{'} = Q_1*R_1$ and $B_i^{'}*B_i=Q_2*R_2$.
Then let $B_{i+1} = Q_1^{'}*B_i*Q_2$. Observe that $B_{i+1}*B_{i+1}^{'}=
R_1*Q_1$ and $B_{i+1}^{'}*B_{i+1} = R_2*Q_2$. Hence, the usual QR iteration
algorithm that was applied on $O_i$ is applied to $B_i*B_i^{'}$ and $B_i^{'}*B_i$ simultaneously. In \cite{bidiagonal},
it is shown that $B_i$ is bidiagonal for all $i$ and converges as $i
\rightarrow \infty$ to a diagonal matrix with the singular values on the
diagonal. Assuming that $B_i$ converges in the second iteration then the
equation can be written as:
$$B_2=Q_1^{'}*B_1*Q_2 = Q_1^{'}*Q_1^{'}*B_0*Q_2*Q_2.$$ 
Where $B=B_0$ is the original bidiagonal matrix. Finally, the singular value
decomposition of matrix B can be represented as:
%S and V are updated at each iteration.  
 $$B=(Q_1^{'}*Q_1^{'})^{-1}*B_2*(Q_2*Q_2)^{-1},$$
which we write as: $$B=U_2*\Sigma*V_2.$$ 
 
  \item {\bf Step 4: Form the singular values and singular vectors of A}

	According to the above steps, we can deduce the following formulas: $A=Q*R$
	(Step 1), $A=Q*U_1*B*V_1$ (Step 2), and $A=Q*U_1*U_2*\Sigma*V_2*V_1$ (Step 3).
	Since the SVD of $A$ is given by $A=U*\Sigma*V^{'}$, the left singular vectors
	$U$ and the right singular vectors $V^{'}$ can be computed by the
	following formulas:
	 
	    $$U=Q*U_1*U_2,\quad V^{'}= V_2*V_1.$$
	
	The columns of $V^{'}$ are the principal components of the input matrix $A$ if
	$A$ is mean-centered (i.e., subtracted by the column mean of $A$)
\end{itemize}

{\bf Time Complexity:}
As shown in Figure \ref{fig:flowchart_svd}, RscaLAPACK PCA implementation uses
routines xGEQRF, xGBBRD, xORGBR. Performance analysis study of such routines
preseneted in \cite{lapack_complexity} shows that the floating point
operations count for each of the above routine is either quadratic or cubic in
the number of dimensions. More specifically, \cite{lapack_complexity} shows that the floating point operations count
performed by the first two routines is  $O(ND^2 + D^3)$.

{\bf Communication Complexity:}
	A disadvantage of this algorithm is that it invokes several routines and each
	routine computes an intermediate output which results in communication
	overhead.
	For large-scale data, communication costs can become a bottleneck for
	scalabilty. The algorithm described in this section consists of three main
	steps that produce intermediate output: QR decomposition and bidiagonalization
	and SVD computation of the bidiagonal matrix. 
	First, QR decomposition of the  $N \times D$ matrix $A$ results in two
	matrices, $N \times d$ matrix $Q$ and $d \times D$ matrix $R$. Therefore, the
	order of intermediate data in this step is $O((N+D)d)$. Second, the
	bidigonalization step of the matrix $R$ results in three matrices: $d \times d$
	matrix $U_1$, $d \times D$ matrix $B$, and $D
	\times D$ matrix $V_1$. Third, the SVD computation of the bidigonal matrix $B$
	results in three matrices of the same dimensions as the ones computed in the bidiagonalization step.
	Hence, the size of intermediate data of each of the previous two steps is
	$O(d^2+Dd+D^2)=O(D^2)$.
	
	To summarize, the total amount of intermediate data generated by the
	algorithm accross the three steps is: $O(max(N+D)d, D^2)$. In this analysis, we
	assume that all elements of the matrices are stored even if some matrices are sparse. This is not practical in real-world applications, however, it provides an upper bound on the
	amount of intermediate data.

\subsection{Singular Value Decomposition for Sparse Matrices}\label{sparse}

SVD implementations optimized for sparse matrices have been explored by some
large-scale machine learning libraries. Mahout provides a Lanczos SVD
algorithm that takes a sparse matrix as an input. The computational complexity
of the algorithm employed by Mahout is $O(Nz^2)$, where $z$ is the number of
non-zero dimensions (out of $D$ dimensions). We refer to this method as
SVD-Lanczos, and it is implemented in popular libraries such as Mahout and
Graphlab. The SVD-Lanczos algorithm, however, is not efficient for performing
PCA for large datasets. This is because the matrix must be mean-centered in
order to obtain the principal components as a result of SVD. Since in many practical
applications the mean of the matrix is not zero, subtracting the mean from a
sparse matrix will substantially decrease its sparsity. In this case, $z$
will approach the full dimensionality $D$, and the cost for computing PCA using
SVD-Lanczos will be $O(ND^2)$, which is prohibitive for large datasets. Graphlab
provides the Lanczos SVD implmentation described in \cite{graphlab_svd}. It is a
slightly different variant of the algorithm implemented in Mahout. The work in
\cite{graphlab_svd} argues that the quality of the solution of the algorithm
implemented in Mahout deteriorates very rapidly when the number of iterations
exceeds a certain value (5-10), and the solution becomes no longer
accurate. Graphlab avoids this problem by supporting a restarted variant of the
Lanczos algorithm in order to enhance the quality of the
output eigenvectors. In this variant, the number of iterations of the Lanczos
algorithm is set to a fixed value, and when this value is reached the algorithm
is re-initiated with a better initial vector. The method for choosing the
enhanced initial vector is described in~\cite{graphlab_svd}. 
%In \cite{graphlab_svd}, it is mentioned that this algorithm is suitable for
% sparse matrices which is not the case in PCA.

\section{Stochastic Singular Value Decomposition (SSVD)}\label{sec:ssvd}

Randomized sampling techniques have recently gained interest in solving
large-scale linear algebra problems. Halko et al. \cite{halko}
discussed randomized methods to compute approximate factorization of matrices
(e.g., SVD). This work proposed a modular framework that computes
approximate SVD in two stages. Stage \rom{1} computes a low dimensional matrix
that approximates the range of the input matrix, which is done via
randomized sampling. Stage \rom{2} uses determenistic techniques to compute the
SVD of the low-dimensional matrix computed in stage \rom{1}. Halko et al.
\cite{halko} describe a set of randomized techniques used in stage \rom{1}
for rapidly constructing low-dimensional approximation matrices. Randomized
methods are often faster and enable computations at large scale compared to the determenistic methods. Furthermore, randomized methods
have the advantage of providing a balance between speed and accuracy.

In the rest of this section, we focus on the analysis of
the Stochastic Singular Value Decomposition (SSVD) algorithm implemented in
Mahout, since the algorithm is designed to scale for large datasets and this is
the focus of our work.
\begin{enumerate}
\item{\bf Stage \rom{1}}

This stage requires an orthonormal matrix $Q$ where $A\approx Q*Q^{'}*A$. $Q$
should contain as few columns as possible and more importantly, $Q$ should approximate the range of $A$ as much as possible (i.e.,
$A$ is as close as possible to $Q*Q^{'}*A$). The algorithm to solve this problem can be formally defined as follows:
Given $N \times D$ matrix $A$, a target dimensionality $d$, and an
oversampling parameter $p$. The algorithm computes an $N \times (d+p)$ matrix
$Q$ that has orthonormal columns and approximates the range of $A$. A matrix $Q$
approximates the range of $A$ if any vector of $Q$ is represented as a linear combination of the vectors of matrix $A$. The
number of extra samples $p$ is small and adds a great deal of accuracy for a small computational cost. As mentioned in \cite{halko}, $p$ is typically set
to 5 or 10, however it can be set to $p = d$ without changing the
asymptotic complexity of the algorithm. In the rest of the discussion, we denote
the term $(d+p)$ by the symbol $l$. The steps for computing the matrix
$Q$ is described as follows:

	\begin{itemize}
	 \item{\bf Step 1: Draw a Gaussian matrix $\Omega$ }
	 
	 Suppose that we are seeking a matrix with $d$ dmensions to approximate the
	 range of Matrix $A$. The idea is to sample $d$ random vectors $
	 \omega^{(i)} ,  i =1,2,\dots d$. For each random vector we compute $z^{(i)}=
	 A*\omega^{(i)}$.
	 Owing to randomness, the set of vectors $\omega ^{(i)}$ are likely to be
	 linear.
	 Hence, each vector $z^{(i)}$ is a linear combination of the vectors of $A$.
	 The matrix comprising all the vectors $z^{(i)}$ can be regarded as an
	 approximate for the range of $A$. The set of vectors $\omega^{(i)}$ are
	 sampled from a standard Gaussian distribution and they form the matrix
	 $\Omega$
	 
	 \item{\bf Step 2: Compute the matrix $Z=A\Omega$}
	 
	 As mentioned in the previous step, the product  $A*\Omega$ results in an
	 $N \times l$ matrix $Z$ that approximates the range of $A$. Hence, $Z$
	 fulfils one of the requirements of the output matrix $Q$. However, the columns
	 of $Z$ are not necessarily orthonormal which is the other requirement on $Q$.
	 Step 4 computes orthornormal vectors using matrix $Z$.
	 
	 \item{\bf Step 3: Perform power iterations}
	 
	 In the cases when the input matrix $A$ has slowly decaying singular values
	 (i.e.  the variance varies slightly accross its dimensions), matrix $Z$
	 does not accurately approximate matrix $A$. The power
	 iteration technique is empoloyed to improve accuracy in such situation. The
	 motivation behind this technique is that singular vectors associated with
	 small singular values interfere with the computation. So the idea is to
	 multiply the matrix $A$ by $A$ and $A^{'}$ alternatly in each iteration. This
	 results in a matrix with higher power that has the same singular vectors as the matrix
	 $A$ but the weight of singular vectors associated with small singular values
	 decreases in the new matrix. The resultant matrix denoted by $(A*A^{'})^t*A$
	 has rapidly decaying singular values. In summary, this step
	 iteratively updates the value of matrix $Z$ according to the below
	 formula:
	 	
	 		$$Z=(A*A^{'})^{'}*A*\Omega,\quad t =1,2,\dots j,$$
	 		  
	 where $j$ is the number of power iterations which is usually
	 a value between 2 and 5.
	 
	 \item{\bf Step 4: Compute QR decomposition of matrix $Z$ }
	 
	 In order to get orthonormal columns from matrix $Z$, the matrix is decomposed
	 into two matrices $Q$ and $R$, such that $Q$ is an orthornoraml matrix and $R$
	 is an upper triangular matrix. Matrix $Q$ computed in this step is a
	 low-dimensional orthornormal matrix and accurately approximates the range of
	 input matrix $A$ which are the requirements on the output matrix from Stage
	 \rom{1} of this algorithm. 
	 
	 QR decompositon of matrix $Z$ is performed by a
	 series Givens rotations described in \cite{givens}.
	 In each iteration, $Z$ is multiplied by a rotation matrix $G_i$. Each rotation
	 converts one of the elements of $Z$ to zero. The series of rotations continues
	 until matrix $Z$ is converted to a triangular matrix $R$, the rotation
	 matrices $G_1,G_2,\ldots G_s$ are multiplied to form the orthogonal matrix
	 $Q$. Assuming that $Z$ is $3\times 3$ matrix. It needs three Givens
	 rotatations to be converted to an upper triangular matrix $R$. The three rotations 
	 $G_1,G_2$ and $G_3$ zeroes the elements $z_{31},z_{21}, and z_{32}$
	 respectively.
	 The product of the rotation matrices forms the matrix $Q^{'}$ (i.e., inverse
	 of matrix $Q$):
	 $$Q^{'}=G_1*G_2*G_3,$$ 
	 and the matrix R is defined as:
	 $$R=G_1*G_2*G_3*Z = Q^{'}*Z.$$ 
	 Therefore, the QR decomposition of $Z$ is defined as: $Z=Q*R$.
	 
   \end{itemize}
   
\item{\bf Stage \rom{2}}

Given the approximate matrix $Q$ computed in stage \rom{1}, $Q$ is used to
compute the SVD of the original matrix $A$. The following steps are performed to
compute the SVD:

\begin{itemize}

	 \item{\bf Step 1: Compute matrices $B=Q^{'}*A$ and $B^{'}*B$}
	
	 In this step, the matrix $B$ is formed by multiplying the transpose
	 of the orthonormal matrix $Q$ with the original matrix $A$. Additionally, a
	 relatively small matrix $B^{'}*B$ is computed in this step.
	 Such matrix has $l \times l $ dimensions so it can be more efficiently stored
	 and used in the following steps.
% 	 Although this step
% 	 might seem redudnant, because $B$ is the same matrix as $R$ that was computed
% 	 in the previous stage, this step is necessary to keep modularity since $Q$ is
% 	 not always formed by QR decomposition. Additionally, SSVD computes $B$ in this
% 	 step to avoid storing a big matrix from the previous stage which is not a
% 	 recommended practice in case of large matrices.
 	 \item{\bf Step 2: Eigenvalue decomposition of $B^{'}*B$}
	 
	 Eigenvalue decomposition is used to factorize matrix $B^{'}*B$ to form the
	 term $B^{'}*B=\widetilde{U}*\Sigma*\widetilde{U}^{'}$, where the columns of
	 $\widetilde{U}$ are the eigenvectors of $B$ and $\Sigma$ is a diagonal matrix
	 whose entries are the eigenvalues of $B$.
	 
	 \item{\bf Step 3: Compute singular vectors of Matrix A}
	 
	 It can be deduced from step 1 that $A=Q*B$.  Multiplying both sides from the
	 right by $B^{'}$ yields the following formula: $A*B^{'}=Q*B*B^{'}$. By
	 applying the decomposition performed in step 2:
	 		
	 		$$A*B^{'}=Q*\widetilde{U}*\Sigma*\widetilde{U}^{'}.$$
	 Multiplying both sides by $(B^{'})^{-1}$:
	 		$$A=Q*\widetilde{U}*\Sigma*\widetilde{U}^{'}*(B^{'})^{-1}.$$
	 Since the SVD of $A$ is defined as $A=U*\Sigma*V^{'}$. Therefore, the left
	 singular vectors $U$ and the right singular vectors $V^{'}$ can be
	 defined as:
	    $$U=Q*\widetilde{U},\quad  V^{'}= \widetilde{U}^{'}*(B^{'})^{-1}.$$ 
	   
\end{itemize}

\end{enumerate}

{\bf Time Complexity:}
Stage \rom{1} generally dominates the cost of Stage \rom{2} in SSVD. Within
Stage \rom{1}, the computational bottleneck is usually the matrix-matrix product
in Step 2. When the matrix $\Omega$ is standard Gaussian, the cost of this
multiplication is $O(NDd)$. However, \cite{halko} shows that ssvd perform such
multiplication in $O(ND\times log(d))$ only. The key idea is to use a structured
random matrix that allows to compute the product in $O(ND \times log(d))$. As described by \cite{halko}, the
simplest example of structured matrices is the subsampled random Fourier
transform (SRFT), An SRFT is an $D \times d$ matrix that has a specific form and properties described in \cite{halko}. When
the random matrix is SRFT, The sample matrix $Z = A*\Omega$ can be computed in
$O(ND \times log(d))$ using subsampled FFT \cite{halko}.

{\bf Communication Complexity:}
The algorithm requires storing several intermediate matrices. Such matrices
include $N \times d$ matrix $Z$, $N \times d$ matrix $Q$, $d \times d$ matrix $R$, $d \times d$ matrix
$B$, $d \times d$ matrix $\widetilde{U}$. The total intermediate data can be calculated as $O(max(Nd, d^2))$.
%  Although, \cite{halko} mentions that Mahout avoids storing big matrices, our
%  experimental results show that the intermediate stor

\section{Probalistic Principal component analysis (PPCA)}\label{sec:ppca}
Probabilistic PCA (PPCA)~\cite{bishop} is a probabilistic approach to computing 
principal components of a dataset. In PPCA, PCA is represented as
a latent variable model that seeks a linear relation between a $D$-dimensional observed data vector $y$ and a
$d$-dimensional latent variable $x$. The model can be described by the equation:
\begin{align*}
\mathbf{y} = \mathbf{C * x} + \mathbf{\mu} + \mathbf{\epsilon},
\end{align*}
where $\mathbf{C}$ is a $D \times d$ transformation matrix (i.e, the columns of
$\mathbf{C}$ are the principal components), $\mathbf{\mu}$ is the vector mean of
$\mathbf{y}$, and $\mathbf{\epsilon}$ is white noise to compensate for errors.
Moreover, \cite{bishop} proposed an Expectation Maximization (EM) algorithm for estimating the principal components iteratively for this latent variable model. PPCA has the following advantages:
\begin{itemize}
  \item {Mathametically proven model for PCA}
  \item {Linear complexity in the number of data points, data dimensions and the
  required number of principal components
\cite{pca_methods} }
  \item {Big data is more susceptible to having missing values. Since PPCA uses EM,
  PCA projections can be obtained when some data values are missing.}
  \item{Multiple PCA models can be combined as a probabilistic
mixture which allows PCA to express complex models}
\end{itemize}

Given a set of $n$-dimensional observed data vectors $Y$, a set of
$d$-dimensional hidden vectors $X$, a set of unknown parameters $C$
and a liklihood function $p(Y | X,C)$. The intuition behind the algorithm is
described in \cite{roweis_empirical} as follows. First, make a random guess for the principal components $C$. Second,
fix the guessed components and project the data vectors $Y$ into it in
order to give the hidden states $X$. Third, fix the values of the hidden
states $X$ and choose $C$ that maximizes $p(Y | X,C)$. The last two steps are
then repeated until convergence. The EM algorithm finds $C$ that maximizes the
liklihood function $p(Y | X,C)$. The main steps are as follows:
	\begin{enumerate}

	\item {\bf E-Step:} Given current parameters $C^{(t)}$, calculate the
	expected value of the log liklihood function $E[log( p(Y | X,C^{(t)} )]$.
	
	\item {\bf M-Step:} Find the parameters that maximizes the expected value
	
	$C^{(t+1)}=E[log(p(Y | X,C^{(t)})]$.

	\item Repeat steps 1 and 2 until convergence.
	
	\end{enumerate}

A standard method for testing whether the algorithm converged or not is
by computing the 1-Norm of the reconstruction error, which is given by: $e = ||Y
- X * C^{-1}||_1$.

{\bf Time Complexity:} 
The analysis presented in~\cite{pca_methods} shows that
EM-based PCA has linear complexity in the number of data points, dimensions and
required principal components; Similar studies in \cite{raiko_linear_complexity}
and \cite{linear_complexity} show that the EM algorithm used to solve PPCA has a
computational complexity of O(NDd) per iteration. Furthermore, an empirical
study conducted by \cite{roweis_empirical} showed that the number of iterations
needed by PPCA to converge is almost constant with changing the dimensionality
D. Therfore the time complexity of one iteration is sufficient to describe the
time complexity of the whole algorithm.

{\bf Communication Complexity:}
The communication complexity of PPCA is bounded by the matrix of hidden
states $X$, which is of size $N \times d$. Therefore, the worst case complexity
of PPCA is $O(Nd)$. However, the work in~\cite{spca} shows that this
complexity can be reduced. In this work, we present our algorithm, {\it sPCA},
which is based on probabilistic PCA. We show that the communication complexity
of the algorithm is $O(Dd)$ without changing any theoritical gurantees provided by the original probabilistic PCA. {\it sPCA} achieves this by redundantly recomputing some large matrices instead of storing them as intermediate data.
Although these redundant computations increase the computational complexity that
has been described previously, such redundant computations are repeated two
times for each iteration. Therefore, the computation complexity becomes
$O(3NDd)$ which does not change the assymptotic complexity of the PPCA
algorithm.

\section{Summary of the analysis} \label{sec:summary}

% The summary of our analysis shows that the methods based on eigenvalue
% decomposition or SVD for dense matrices are computationally intensive as they require $O(NDmin(N,D))$ flops. 
% 
% Similarly, SVD-based algorithm that is optimized for sparse matrices is not
% efficient when applied to PCA. In order to compute PCA, such algorithms perform
% SVD on a mean-centered matrix. Since the mean is usually non-zero, the input
% matrix is filled with non-zero elements and the complexity becomes as
% high as SVD for dense matrices; $O(ND^2)$.
% 
% On the other hand, random sampling techniques for computing SVD such as
% Stochastic SVD (SSVD) has computational complexity of $O(NDd)$ and sometimes,
% $O(ND \times log(d))$. Such algorithms proved to be faster and more scalable
% than classical methods for computing SVD and its worst case complexity is similar to
% that of PPCA. Hence, the two latter methods are potential candidates for
% performing PCA for large datasets. However, flops is not the only bottleneck.
% With the wide access of parallel computing clusters, communication overhead emerges as a real bottleneck. In such environments, communication costs
% often dominates the parallel computation costs. Even though the time complexity
% of stochastic SVD can be handled by employing more computing nodes, our analysis
% shows that it suffers from substantial communication complexity; $O(Nd)$

\begin{table*}[tp]
\centering
\begin{tabular}{|c|c|c|c|}
\hline
	{\bf Method to Compute PCA} & {\bf Time Complexity} & {\bf Communication Complexity} & {\bf Example Libraries} \\
\hline			
	Eigen decomp. of covariance matrix & $O(N D \times \min(N,D) + D^3)$ &
	$O(D^2)$ & \vtop{\hbox{\strut MLlib-PCA (Spark),}\hbox{\strut RScaLAPACK}} \\
\hline
	SVD-Bidiag \cite{demmel_kahan} & $O(ND^2 + D^3)$ & $O(\max((N+D)d, D^2))$  &
	RScaLAPACK   \\
 \hline
% 	SVD-Lanczos \cite{graphlab_svd} & $O(N z^2)$ &  & Mahout, GraphLab  \\
%\hline
Stochastic SVD (SSVD) \cite{halko} & $O(N D d)$ & $O(\max (Nd, d^2))$ &
Mahout-PCA (MapReduce)  \\
\hline
Probabilistic PCA (PPCA) \cite{ppca}  & $O(N D d)$ & $O(Dd)$ & sPCA~\cite{spca}
\\
\hline
\end{tabular}
\caption{Comparison of different methods for computing PCA of an $N \times D$ matrix to produce $d$ principal components.}
\label{tbl:analysis_summary}
\end{table*}

The summary of our analysis is shown in Table~\ref{tbl:analysis_summary}. As the table shows, the time complexities of the top two methods (eigen decomposition 
of covariance matrix and SVD of bi-diagonalized matrix)
%are a function of $N$ (number of data points) multiplied by $D^2$ 
%(number of dimensions of each data point), which is quite high for many datasets with a large number of dimensions. 
are approximately cubic in terms of the dimensions of the input matrix, assuming 
$N \approx D$. Such high time complexities prevent these methods from 
scaling to large datasets. Even if $D < N$, the time complexity is  
a function of $N$ (number of data points) multiplied by $D^2$ 
(number of dimensions of each data point), which is still quite high 
for many datasets with large number of dimensions.
In addition, the communication complexities of these two methods are also quite
high, especially for high dimensional datasets.
Therefore, even if there are enough computing nodes to handle the high
computational costs, the communication costs can still hinder the scalability of these two methods. 

The last two methods in Table~\ref{tbl:analysis_summary} (stochastic SVD and 
probabilistic PCA) have a more efficient time complexity of $O(ND)$, assuming 
that $d$ is a relatively small constant, which is typically the case in many
real applications. Thus, these two approaches are potential candidates for performing PCA for 
large datasets. Our analysis and experimental evaluation~\cite{spca}, however,
reveal that even though the time complexity of stochastic SVD can be handled by employing more computing nodes,
it can suffer from high communication complexity. For example, our
experiments show that the high communications complexity of Mahout-PCA (which
uses SSVD) prevents it from processing datasets in the order of tens of GBs. Therefore, based on our analysis, the most
promising PCA approach for large datasets is the probabilistic PCA.

%For example, our experiments show that the high communications complexity of
% Mahout-PCA (which uses SSVD) prevents it from processing datasets in the order of tens of GBs.

\section{Conclusion} \label{sec:conc}

In this report, we analyzed different methods for computing the principal
components of a given dataset, which is referred to as principal component
analysis (PCA). The analysis showed that both the computational complexity as
well as the communications complexity are important aspects for processing
large-scale datasets on distributed platforms. Our analysis indicated that the
two methods: eigen decomposition 
of covariance matrix and SVD of bi-diagonalized matrix are omputationally
intensive as their time complexities are either cubic in terms of the dimensions
of the input matrix or a function of $N$ (number of data points) multiplied by
$D^2$ (number of dimensions of each data point), which is quite high for many datasets. On the other hand, the analysis showed that
Stocahstic SVD (SSVD) and Probablistic PCA are two potential candidates for performing PCA on large datasets, since they have
the best computational complexity. However, SSVD suffers from high communication complexity. 
Therefore, the most promising PCA approach for large
datasets is the probabilistic PCA. In our recent work~\cite{spca}, we have
designed a distributed and scalable principal component analysis algorithm
called {\it sPCA} and it is based on probabilistic principal component analysis.
sPCA substantially outperforms current PCA algorithms implemented in
Mahout/MapReduce and MLlib/Spark.

% \input{background}
% \input{design}
% \input{impl}
% \input{eval}
% \input{related}
% \input{conclusion}

%\bibliographystyle{abbrv}
%\bibliography{paperlist}

\end{document}